\documentclass[]{spie}  

\usepackage{amsmath,amsfonts,amssymb}
\usepackage{graphicx}
\usepackage[colorlinks=true, allcolors=blue]{hyperref}

\title{Sub-diffraction-limited coronagraphic imaging\\ with nano-printed PIAACMC phase masks}

\author[a]{Elena Tonucci}
\author[a,b]{Sebastiaan Y. Haffert}
\author[b]{Warren B. Foster}
\author[b]{Jared R. Males}
\author[b,c,d,e]{Olivier Guyon}
\author[b]{Laird M. Close}
\author[b]{Kyle van Gorkom}
\author[f]{Alexander D. Hedglen}
\author[b]{Parker T. Johnson}
\author[b]{Maggie Y. Kautz}
\author[c]{Jay K. Kueny}
\author[a]{Rico Landman}
\author[b]{Jialin Li}
\author[c]{Joshua Liberman}
\author[g]{Joseph D. Long}
\author[b,d]{Miles Lucas}
\author[c]{Jennifer Lumbres}
\author[a]{Matthijs Mars}
\author[c]{Eden A. McEwen}
\author[h]{Avalon McLeod}
\author[c]{Tiffany Nguyen}
\author[i]{Logan A. Pearce}
\author[a]{María Eugenia Redondo González}
\author[j]{Lauren Schatz}
\author[c]{Katie Twitchell}
\affil[a]{Leiden Observatory, Leiden University, PO Box 9513, 2300 RA, Leiden, The Netherlands}
\affil[b]{Steward Observatory, The University of Arizona, 933 North Cherry Avenue, Tucson, Arizona}
\affil[c]{Wyant College of Optical Sciences, The University of Arizona, 1630 E University Blvd, Tucson, Arizona}
\affil[d]{Subaru Telescope, National Observatory of Japan, National Institutes of Natural Sciences, 650 N. A'ohoku Place, Hilo, Hawai'i}
\affil[e]{Astrobiology Center, National Institutes of Natural Sciences, 2-21-1 Osawa, Mitaka, Tokyo, Japan}
\affil[f]{Northrop Grumman Corporation, 600 South Hicks Road, Rolling Meadows, Illinois}
\affil[g]{Center for Computational Astrophysics, Flatiron Institute, 162 5th Avenue, New York, New York}
\affil[h]{Draper Laboratory, 555 Technology Square, Cambridge, Massachusetts}
\affil[i]{Department of Astronomy, University of Michigan, Ann Arbor, MI 48109, USA}
\affil[j]{Starfire Optical Range, Kirtland Air Force Base, Albuquerque, New Mexico}

\authorinfo{Further author information: (Send correspondence to E.T.)\\E.T.: E-mail: tonucci@strw.leidenuniv.nl}

\begin{document} 
\maketitle

\begin{abstract}
Imaging Earth-like exoplanets in the habitable zone of their host star is among the main science objectives of future ground-based and space-based observatories. However, the extreme contrast and small separations needed to image such planets cannot be reached with current technology. The Phase-Induced Amplitude Apodization Complex Mask Coronagraph (PIAACMC) is a promising coronagraph to reach this goal. The PIAACMC uses a set of aspheric lenses to apodize the entrance pupil without throughput losses and a phase-shifting focal plane mask for starlight suppression. These allow us to maintain high throughput and achieve a small inner-working angle (IWA), unlocking the capability to observe exoplanets at the diffraction limit. The masks are manufactured in-house at Leiden University with Nanoscribe, a micro-3D-printer that uses two-photon polymerization to achieve sub-micron precision in height. We present the first scientific results with a focal plane mask for the PIAACMC on the Magellan Adaptive Optics eXtreme (MagAO-X) instrument for the 6.5-meter Magellan Clay telescope at Las Campanas Observatory, Chile. We show laboratory and on-sky contrast curves with a broadband z' filter centered at 908 nm with a 14\% bandwidth. We use the PIAACMC to detect binary companions at separations $\sim$0.8-5 $\lambda$/D ($\sim$23-144 mas). This demonstrates the PIAACMC’s capability to observe at the diffraction limit and below, with a sub-$\lambda$/D IWA. Future work includes exploring new mask designs to improve the contrast in broadband light and performing active focal plane wavefront sensing and control.
\end{abstract}


\section{INTRODUCTION}
\label{sec:intro}
As we enter the era of Extremely Large Telescopes (ELTs\cite{ELT,GMT}), developing technologies to hunt for temperate, Earth-sized exoplanets in the habitable zone of their host star is of outmost importance. Among the techniques that will make it possible to study these exotic worlds and possibly discover life, direct imaging is one of the most promising thanks to its ability to spatially separate stellar and planetary light\cite{HWO_biosignatures}. This is combined with coronagraphy, which suppresses on-axis starlight while leaving off-axis planetary light unchanged. To observe terrestrial planets however, it is necessary to develop coronagraphs that are able to reach deep contrasts and possess a small IWA and a high throughput\cite{coro_review,tonucci_PIAACMC}.

In this context, the PIAACMC\cite{PIAACMC_og} is a promising option thanks to its high throughput that approaches the physical theoretical limit, and its small IWA below the diffraction limit. This is made possible thanks to: 1) phase-induced amplitude apodization (PIAA) aspheric lenses that apodize the entrance pupil without losses, and 2) a phase-shifting focal plane mask which does not reject or block starlight, but instead exploits interference to suppress it. Moreover, the PIAACMC's performance is not affected by light contamination caused by diffractive elements in the telescope aperture\cite{Guyon_apertures}, making it an ideal candidate for segmented and on-axis telescope designs.

Since its conceptualization, the PIAACMC has been tested with different implementations, primarily in laboratory conditions, for example with the High Contrast Imaging Testbed (HCIT) at the Jet Propulsion Laboratory in Pasadena, where they demonstrated a contrast of 3$\times$10\textsuperscript{-8} at 3-9 $\lambda$/D with 2\% bandwidth, and a contrast of 1.9$\times$10\textsuperscript{-8} at 3.5-8 $\lambda$/D with 10\% bandwidth\cite{HCIT1,HCIT2}. The PIAACMC is also implemented on the Segmented pupil Experiment for Exoplanet Detection (SPEED) testbed\cite{SPEED} at the Lagrange Laboratory in Nice, and on the Subaru Coronagraphic Extreme Adaptive Optics system (SCExAO)\cite{Lozi_paper} at the Subaru telescope in Hawaii. With SCExAO, a preliminary demonstration was performed in H band in lab conditions\cite{Lozi_poster} and on-sky\cite{Knight_thesis}. The first full on-sky performance characterization of PIAACMCs was performed in our previous work Tonucci et al. 2026\cite{tonucci_PIAACMC}, with the MagAO-X\cite{MagAO-X} instrument on the 6.5-meter Magellan Clay telescope at Las Campanas Observatory, in Chile, at sub-micron near-infrared wavelengths, approaching the visible. 
Our main findings included sub-diffraction IWAs of about 0.75 $\lambda$/D in two filters, a narrowband filter centered at 895 nm with a 3\% bandwidth, and a broadband z' filter centered at 908 nm with a 14\% bandwidth. It also served as a PIAA alignment demonstration on-sky. As for the contrast performance, the raw contrast values reached are given as an average in the region within 1 and 5 $\lambda$/D, a range of small separations where the PIAACMC can still observe. With the internal source, we reached a contrast of about 1.6$\times$10\textsuperscript{-3} with the 875 filter and 1.3$\times$10\textsuperscript{-3} with the z’ filter. The main limitations were the focal plane masks' manufacturing errors, residual jitter, and uncorrected quasi-static speckles in MagAO-X. On-sky, we reached a contrast of about 1.4$\times$10\textsuperscript{-2} with the 875 filter and 7.8$\times$10\textsuperscript{-3} with the z’ filter. These were additionally limited by atmospheric conditions and imperfectly corrected aberrations.

In this work, we show the performance of a new PIAACMC phase mask with the broadband z' filter, making this coronagraph fully commissioned on MagAO-X. We also show observations of tight stellar binaries, demonstrating that the PIAACMC enables sub-diffraction-limited coronagraphic imaging. In Section \ref{sec:manuf} we discuss the new design and manufacturing results of the PIAACMC's focal plane mask. In Section \ref{sec:results} we show the performance of the new mask with the internal source of the instrument and on-sky, and demonstrate the PIAACMC's ability to perform sub-diffraction imaging with observations of close-in binaries. Finally, in Section \ref{sec:conclusions} we summarize our findings and discuss future work.

\section{DESIGN AND MANUFACTURING}\label{sec:manuf}
For this work, we designed and manufactured a focal plane mask that phase-shifts light as it propagates through it. The other components of the PIAACMC on MagAO-X were left unchanged. For a detailed explanation of the PIAACMC's layout and its components in MagAO-X, see the on-sky demonstration paper Tonucci et al. 2026\cite{tonucci_PIAACMC}. We only explored designs for the broadband z' filter on MagAO-X, at 908 nm $\pm$ 65 nm ($\sim$14\%), to have a big enough band to collect as many photons as possible during science observations. Given the inherent chromaticity of a phase shifting mask, designing masks for narrower filters should provide similar or even better results. The diameter of the mask was set at 1.5 $\lambda$/D, which was found to be the optimal value during a study on mask size versus achievable performance. With MagAO-X's F number being $f/69$, this corresponds to a physical diameter of $\sim$94 $\mu$m. The mask was parameterized into concentric rings each with a different physical height. To keep the computation times reasonable, the number of rings was limited to ten, as simulations revealed no significant performance increase with a higher number.

The mask was designed with a custom code developed in-house that optimizes on-axis light suppression. To build up from previous work, we improved the design by introducing tolerance to surface deviations during the optimization process, using error values we expected from previous manufacturing tests. We used the Newton-CG minimization method implemented in SciPy\cite{scipy} while supplying gradients that were calculated analytically with reverse-mode algorithmic differentiation relations\cite{diff_relations}. The MagAO-X system was simulated entirely with the package High Contrast Imaging for Python (HCIPy)\cite{hcipy}.

The manufacturing of the focal plane masks was carried out in-house in Leiden University with the Nanoscribe Photonic Professional GT machine, which allows micro-fabrication through additive manufacturing based on two-photon polymerization. The prints were made on a 1mm-thick ultraviolet fused silica uncoated substrate. The material used is the photo-resistant resin IP-Dip, and we used the 63x immersion objective, to achieve the highest resolution possible. For more information on the exact printing process and parameters, which were fundamentally unchanged, see our previous work\cite{tonucci_PIAACMC}.

\begin{figure}[h]
    \centering
    \includegraphics[width=0.5\linewidth]{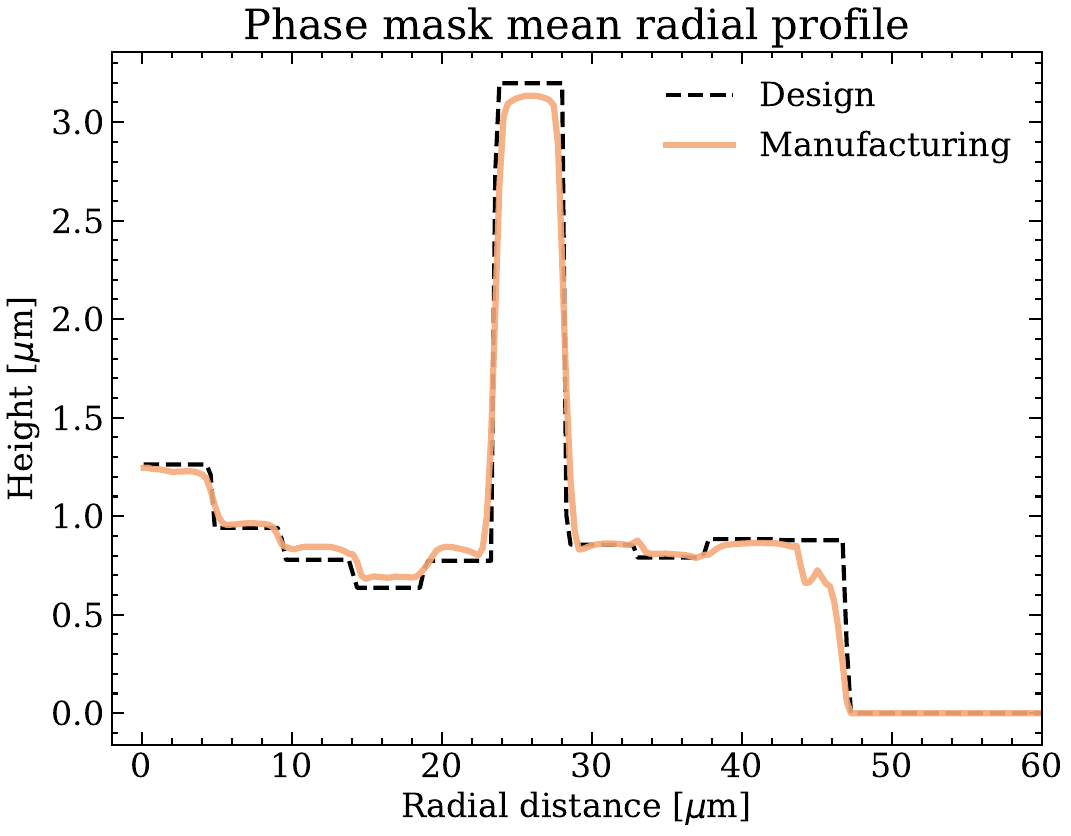}
    \caption{Mean radial profile of designed (dashed black line) and most closely manufactured (colored solid line) focal plane masks.}
    \label{fig:mask_profile}
\end{figure}

We printed six of the same masks on a same substrate with different offsets from the interface, to mitigate possible machine miscalibrations and maximize our chances to get good prints. The masks were spaced enough between each other to avoid interference between them during observations. We then measured the height profile of each mask with a 3D laser scanning microscope at the Space Research Organisation Netherlands (SRON) and chose the mask that provides the best on-axis light suppression to use for testing and observations. Figure \ref{fig:mask_profile} shows the the mean radial profiles of the most closely manufactured mask with respect to the design. We measured surface deviations of order $\sim$10 nm, with respect to the surface deviations of order $\sim$100 nm found in our previous work. We are currently unsure about the repeatability of this result, and will follow the same approach of printing different masks for redundancy in the future.

\begin{figure}[h]
    \centering
    \includegraphics[width=0.5\linewidth]{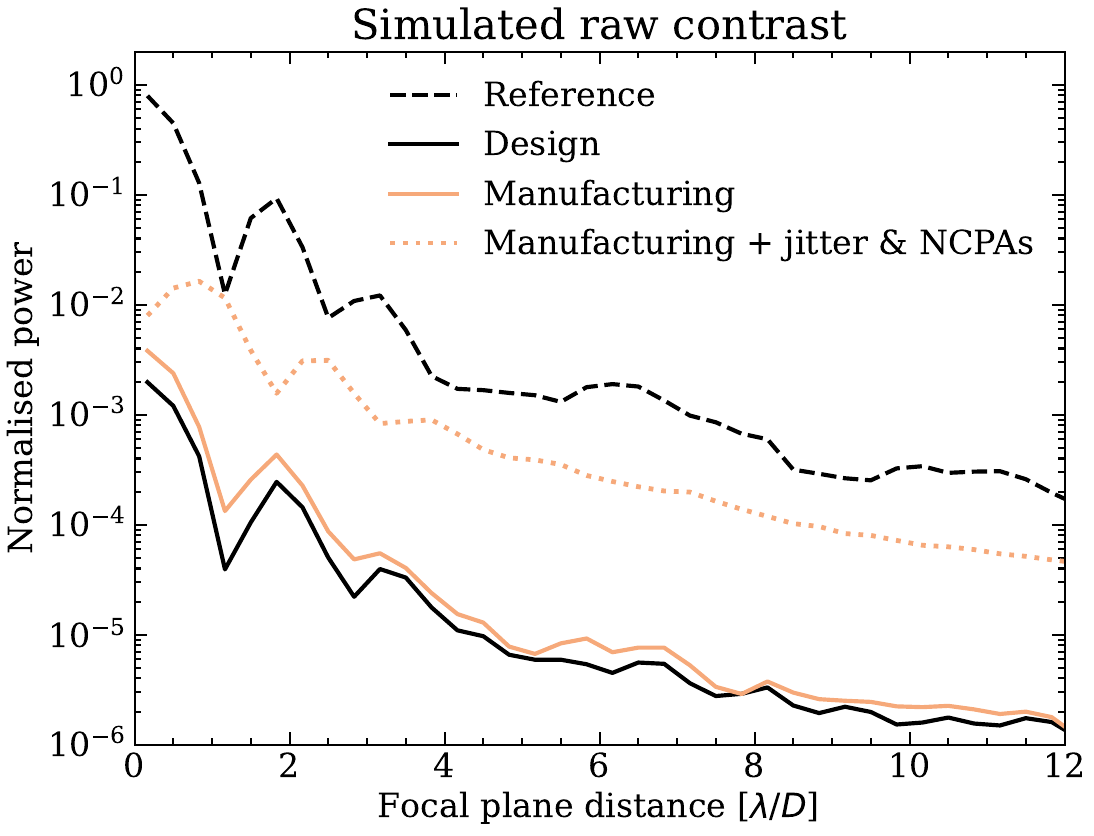}
    \caption{Simulated raw contrast curves as a mean radial profile with the designed mask in ideal conditions (solid black line), with the manufactured mask in ideal conditions (solid colored line), and with the manufactured mask in perturbed conditions including typical values of residual jitter and uncorrected NCPAs in MagAO-X (dotted colored line). The dashed black line shows the simulated non-coronagraphic point spread function (PSF) as a reference (only the mask is removed from the simulation, the rest is left unchanged).}
    \label{fig:sim_contrast}
\end{figure}

Figure \ref{fig:sim_contrast} shows the simulated raw contrast reached with the designed mask and with the most closely manufactured mask in perfect conditions (no aberrations, other optical components are perfect). To simulate the manufactured mask, the measured height map from the laser scanning microscope was used in the end-to-end physical optics model. Additionally, we show the effect of typical MagAO-X residual jitter of $\sim$7 mas, and residual uncorrected non-common path aberrations (NCPAs) of $\sim$2$\times$10\textsuperscript{-3} on the performance with the manufactured mask. The PIAACMC is highly sensitive to tip-tilt errors and these are currently the limiting factors in contrast performance in MagAO-X.

Again, we quote values of contrast within the range of small separations from 1 to 5 $\lambda$/D, since the PIAACMC's strongest point is the capacity to observe at small separations. At larger separations, clearly, the contrast will be even better. The simulated raw contrast is about 4.4$\times$10\textsuperscript{-5} with the designed mask (about a factor two better than our previous work), and about 7.6$\times$10\textsuperscript{-5} with the manufactured mask. The difference between performance with designed and manufactured masks is only about a factor two, a result made possible by the low surface errors we achieved and our new optimization process that provides a mask design more robust to printing errors. Finally, the performance we expect to measure in laboratory experiments is the one with perturbations included in the simulation. In this case, we reach a contrast of about 1.6$\times$10\textsuperscript{-3}, which is almost exactly the same result of our previous work. This clearly shows that the design and manufacturing of the mask is currently not the limiting factor in performance in MagAO-X. While it is still very important to improve our mask design optimizations and to test repeatability of the manufacturing process, the most high-priority improvement for the system is vibration control.

\section{RESULTS}\label{sec:results}
\subsection{Laboratory and on-sky performance}
The throughput curve and IWA of the PIAACMC on MagAO-X is shown in our previous work, Tonucci et al. 2026\cite{tonucci_PIAACMC}. Even if its focal plane mask is now new, the throughput performance remains unchanged and the IWA is about 0.76 $\lambda$/D.

\begin{figure}[h]
    \centering
    \includegraphics[width=0.5\linewidth]{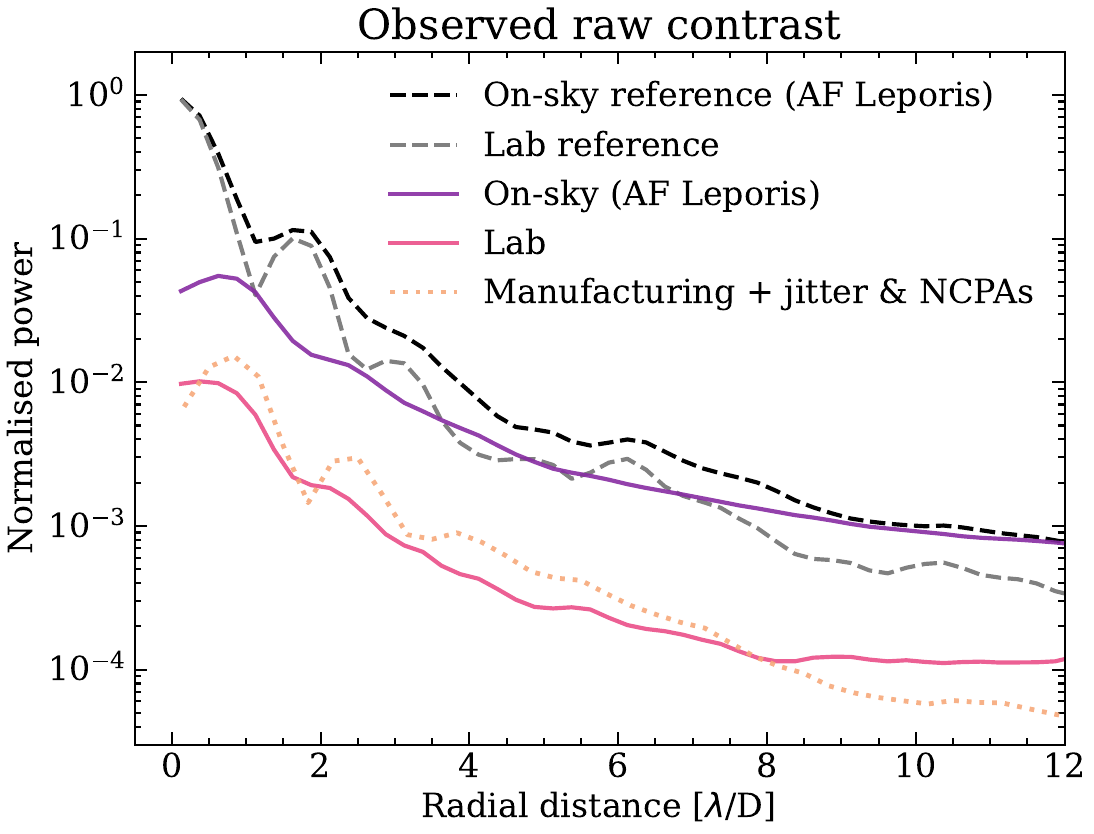}
    \caption{Measured raw contrast curves as a mean radial profile with the internal source of the instrument (solid purple line) and of on-sky observations of AF Leporis (solid fuchsia line). The simulated raw contrast with the manufactured mask in perturbed conditions including typical values of residual jitter and uncorrected NCPAs in MagAO-X from the previous figure is still shown as a reference (dotted line). The dashed gray line and the dashed black line show the measured non-coronagraphic PSF with the internal source and on-sky, respectively (only the mask is removed from the beam, the rest is left unchanged).}
    \label{fig:obs_contrast}
\end{figure}

To study the contrast performance, we take measurements with the internal source of the instrument and on-sky observations of the F-type star AF Leporis. We observed this and all the targets in the next section during MagAO-X's 2025B observing run. We observed AF Leporis specifically on the 30\textsuperscript{th} of November 2026 starting at 05:22 UTC, experiencing variable seeing conditions between 0.65 and 0.9 arcsec. The adaptive optics loop was closed on all 1563 modes. Figure \ref{fig:obs_contrast} shows the measured contrast curves.

We quote contrast values as the average raw contrast in the region within 1 and 5 $\lambda$/D. We measure a contrast of about 9.6$\times$10\textsuperscript{-4} with the internal source and of about an order of magnitude worse, 9.3$\times$10\textsuperscript{-3}, on-sky. These results are very close to the measurements of our previous work. Moreover, the laboratory results agree extremely well with the performance we expected because of residual jitter and uncorrected NCPAs, that limit our performance in laboratory conditions. On-sky, the performance is additionally strongly limited by imperfectly corrected atmospheric aberrations.

\subsection{On-sky observations}
\begin{figure}[h]
    \centering
    \includegraphics[width=0.9\linewidth]{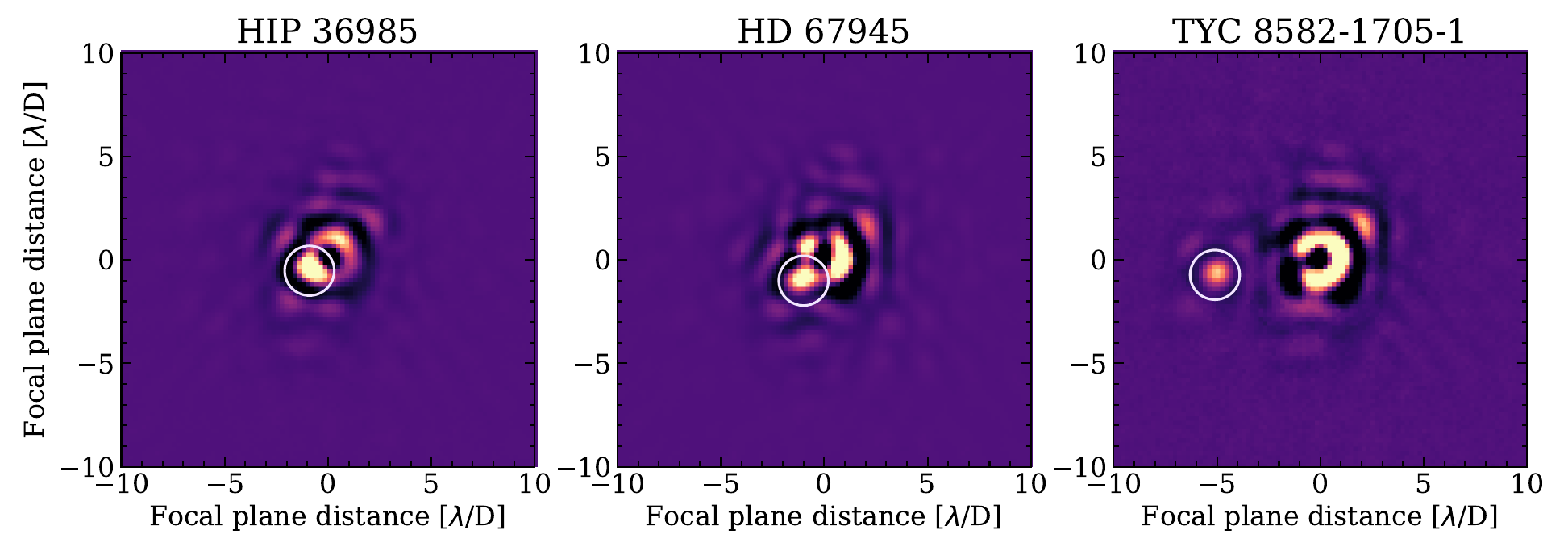}
    \caption{On-sky images of the binary stars HIP\,36985 AB (left panel, $\sim$1 $\lambda$/D separation), HD\,67945 AB (central panel, $\sim$1.5 $\lambda$/D separation), and TYC\,8582-1705-1 AB (right panel, $\sim$5 $\lambda$/D separation). All panels are shown in the same linear scale. The images were high-pass filtered to remove residual atmospheric structures.}
    \label{fig:binaries}
\end{figure}

First, we observed some close-in binaries from the SpHere INfrared survey for Exoplanet (SHINE). At the time of our target selection, they had published results from 78 multiple star systems of which 56 were new discoveries\cite{binaries_SHINE}. The full sample characterization is now also available\cite{SHINE_full}. We selected a few stars from the SHINE survey making sure they would be observable during our observing run, and gave priority to those expected at small separations. This is because we wanted to demonstrate the observing capabilities of the PIAACMC at the diffraction limit. Figure \ref{fig:binaries} shows detections of three binary systems, all observed on the 30th of November starting at 08:12 UTC, with a seeing around 0.6 arcsec.

\begin{itemize}
    \item \textbf{HIP\,36985 AB.} An M1 red dwarf primary with a mid-M dwarf companion for which orbital solutions and dynamical masses are well determined\cite{HIP}. The separation during our observation is about 29 mas (1 $\lambda$/D), and the contrast is roughly 6.7$\times$10\textsuperscript{-2}.
    \item \textbf{HD\,67945 AB.} An F0 primary with a newly detected companion (likely a low-mass star) by the SHINE survey. The exact spectral type, dynamical mass, and orbit of the companion, as well as the system's age, are not yet constrained. The separation during our observation is about 43 mas (1.5 $\lambda$/D), and the contrast is roughly 5.5$\times$10\textsuperscript{-2}.
    \item \textbf{TYC\,8582-1705-1 AB.} A K or G dwarf primary with a newly detected very low-mass star companion by the SHINE survey. They obtained some constraints on the orbit with the Monte Carlo method, but spectral type, dynamical mass, orbit, and age are not yet well constrained. The separation during our observation is about 144 mas (5 $\lambda$/D), and the contrast is roughly 2.8$\times$10\textsuperscript{-2}.
\end{itemize}

\begin{figure}[h]
    \centering
    \includegraphics[width=0.9\linewidth]{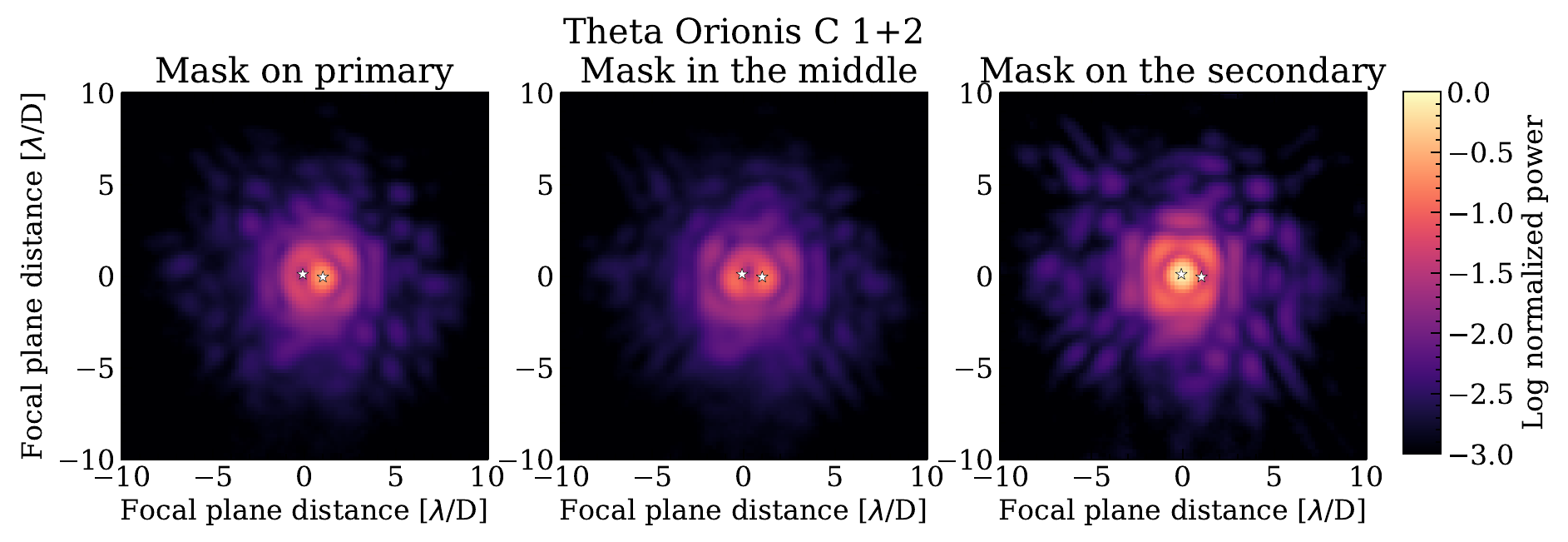}
    \caption{On-sky images of Theta Orionis C1 and C2, at $\sim$0.8 $\lambda$/D separation. All panels are shown in the same logarithmic scale. The left panel shows the focal plane mask aligned on the brighter star on the left, while the fainter star on the right is visible. The central panel shows the mask in the center of the two stars. The right panel shows the mask aligned on the fainter star on the right, while the brighter star on the left is visible.}
    \label{fig:trapezium}
\end{figure}

Finally, during the 4th of December starting at 08:09 UTC, we observed Theta Orionis C, a triple star system. The primary, C1, is a spectroscopic binary star with an angular separation of $\sim$2.5 mas\cite{trapezium_C1ab}, while the companion star, C2, orbits at an angular separation of 20 to 30 mas from C1\cite{trapezium_C12}. C1 and C2 are both O-type stars. Figure \ref{fig:trapezium} shows the detection of the two main components of Theta Orionis C. The separation during our observation is currently estimated at about 23 mas (0.8 $\lambda$/D), and the contrast is roughly 1.7$\times$10\textsuperscript{-1}. We scanned the focal plane mask from left to right using picomotors actuating the mask's filter wheel on MagAO-X, to show the different components of the system. In the left panel, the mask is centered on the brighter star located on the left, showing the PSF of the fainter star on the right. In the central panel, the mask was moved in between the two stars, effectively showing some light from both of them. Since the separation is very small, the PSFs are not clearly visible here, rather, we can see two distorted coronagraphic images leaking through the phase mask on both sides. Finally, in the right panel, the mask was moved and aligned on the fainter star located on the right, showing the PSF of the brighter star on the left. This observation demonstrates the PIAACMC's capability of observing not only at but even below the diffraction limit.

\section{CONCLUSIONS}\label{sec:conclusions}
We have shown coronagraphic observations of binary companions at the diffraction limit and below with a z' broadband filter, using the PIAACMC on MagAO-X. With respect to our previous work, the design of the phase mask was improved in robustness against typical Nanoscribe manufacturing errors. We reached surface deviations of order $\sim$10 nm in height, allowing an on-axis suppression very close to the design, in perfect conditions. The average raw contrast we achieved in the region within 1 and 5 $\lambda$/D, very interesting for close-in companions, is about 9.6$\times$10\textsuperscript{-4} with the internal source and about one order of magnitude worse, 9.3$\times$10\textsuperscript{-3} on-sky. Such values are well aligned with our previous work, showing that the PIAACMC is not currently limited by the design or manufacturing of its phase mask, but by uncorrected NCPAs and residual jitter. The PIAACMC is, in fact, very sensitive to tip/tilt errors. For our first science observations, we showed detections of close-in binary companions including HIP\,36985 AB, HD\,67945 AB, TYC\,8582-1705-1 AB, and Theta Orionis C1 and C2. While TYC\,8582-1705-1 AB are at a larger separation of 5 $\lambda$/D, HIP\,36985 AB and HD\,67945 AB are companions at the diffraction limit (1 and 1.5 $\lambda$/D, respectively). Finally, Theta Orionis C1 and C2 are estimated at a separation of only 0.8 $\lambda$/D, confirming the PIAACMC's capabilities of imaging companions at the diffraction limit and even below.

Future work will focus on three fronts: The phase mask design, the MagAO-X upgrades, and the active focal plane wavefront sensing and control. First, we will explore new mask concepts aimed at improving the achievable contrast in broadband light, which will require new optimizations and fine-tuning of the manufacturing process. On the MagAO-X side, the highest priority improvement is the management of bench turbulence and of telescope vibrations with predictive control, to reduce the residual jitter. Finally, we want to integrate active focal plane wavefront sensing and control with the PIAACMC, to reduce NCPAs and quasi-static speckles. We plan to perform the sensing part with pair-wise probing\cite{PWP} first, and then with the Self-Coherent Camera\cite{SCC} through the Fast Atmospheric SCC Technique\cite{FAST}. The control will be performed through (implicit) Electric Field Conjugation\cite{EFC,iEFC,iEFC_onsky}. With our work at existing facilities, we are demonstrating that the PIAACMC is a promising candidate for future ground-based and space-based observatories. The PIAACMC will enable observations of a whole new population of companions located at the diffraction limit, a region inaccessible to other coronagraphs.


\acknowledgments 
This paper includes data gathered with the 6.5 metre Magellan Telescopes located at Las Campanas Observatory, Chile. The MagAO-X phase II project acknowledges generous support from the Heising-Simons Foundation. We are very grateful for support from the NSF MRI Award \#1625441 (MagAO-X). MagAO-X uses the CACAO software package, which is supported by NSF Award \#2410616. SYH acknowledges support from NWO Award 184.036.004. JL and SYH acknowledge support from NASA APRA award 80NSSC24K0288.

\bibliography{report} 
\bibliographystyle{spiebib} 

\end{document}